\documentclass[12pt,preprint]{aastex}

\def\psfig#1{\relax}

\long\def\Ignore#1{\relax}

\def\today{\count99=\day
           \ifnum\count99>20 \count98=\day
                             \divide\count98 by 10
                             \multiply\count98 by 10
                             \advance\count99 by -\count98 \fi
           \number\day\ifcase\count99 th\or st\or nd\or rd\else th\fi
           ~\ifcase\month none\or January\or February\or March\or April\or
                  May\or June\or July\or August\or September\or October\or
                  November\or December\fi
           ~\number\year}

\shorttitle{Fabry-Perot Spectroscopy of NGC 7079}
\shortauthors{Debattista \& Williams}

\begin{document}

\def\om{\Omega_p}
\def\len{a_B}
\def\lag{D_L}
\def\vpd{{\cal R}}
\def\n7{NGC 7079}
\def\degrees{^\circ}
\def\arcmin{^\prime}
\def\arcsec{^{\prime\prime}}
\def\kin{{\cal V}}
\def\pin{{\cal X}}
\def\padisk{PA$_{\rm disk}$}
\def\dpa{\delta_{\rm PA}}
\def\kms{$\mathrm {km}\ \mathrm s^{-1}$}
\def\kmsa{$\mathrm {km}\ \mathrm {s}^{-1} \mathrm{arcsec}^{-1}$}
\def\kmsm{$\mathrm {km}\ \mathrm s^{-1} \mathrm{Mpc}^{-1}$}
\def\kmsk{$\mathrm {km}\ \mathrm s^{-1} \mathrm{kpc}^{-1}$}
\def\mas{mag arcsec$^{-2}$}   
\def\iraf{IRAF}
\def\etal{{\it et al.}}
\def\eg{{\it e.g.}}
\def\etc{{\it etc.}}
\def\ie{{\it i.e.}}
\def\cf{{\it cf.}}
\def\spose#1{\hbox to 0pt{#1\hss}}
\def\gtsim{\mathrel{\spose{\lower.5ex \hbox{$\mathchar"218$}}
     \raise.4ex\hbox{$\mathchar"13E$}}}
\def\ltsim{\mathrel{\spose{\lower.5ex\hbox{$\mathchar"218$}}
     \raise.4ex\hbox{$\mathchar"13C$}}}
\def\nd{\nodata}

\title{Fabry-Perot Absorption-Line Spectroscopy of \n7: \\ Kinematics and Bar
Pattern Speed\footnote{Based in part on observations carried out at the European Southern Observatory (Prop. No. B-0329)}}
\author{Victor P.\ Debattista\altaffilmark{2,3}}
\affil{Institut f\"ur Astronomie, ETH  H\"onggerberg, HPF G4.2, CH-8093, 
Z\"urich, Switzerland, debattis@phys.ethz.ch}
\author{T.\ B.\ Williams\altaffilmark{3}}
\affil{Department of Physics and Astronomy, Rutgers, The State 
University of New Jersey, 136~Frelinghuysen Road, Piscataway, NJ 08854-8019,
williams@physics.rutgers.edu}
\altaffiltext{2}{Astronomisches Institut, Universit\"at 
Basel, Venusstrasse 7, CH-4102 Binningen, Switzerland}
\altaffiltext{3}{Visiting Astronomer, Cerro Tololo Inter-American Observatory 
(CTIO).}

\begin{abstract}
We present Fabry-Perot absorption-line spectroscopy of the SB0 galaxy \n7.
This is the first use of Fabry-Perot techniques to measure the
two-dimensional stellar kinematics of an early-type disk galaxy.  
We scan the infrared CaII line using the Rutgers Fabry-Perot (RFP), to obtain 
kinematic data extending to $I$-band surface brightness 
$\mu_I \simeq 21$ \mas, in a field of radius $\sim 40\arcsec$.  
The kinematic data, consisting of line-of-sight velocities and 
velocity dispersions, are in good agreement with data obtained along
the major axis of the disk with standard slit spectroscopy.  
Comparison of the exposure times required for slit and RFP spectroscopy 
to reach the same limiting magnitude shows that the RFP is significantly
more efficient for mapping absorption-line galaxy kinematics.
We use the velocity data, together with our own deep broad-band photometry,
to measure the bar pattern speed, $\om$, of \n7\ with the model-independent
Tremaine-Weinberg (TW) method.  
We find $\om = 8.4 \pm 0.2$ \kmsa; this is the best-constrained pattern 
speed ever measured for a bar using the TW method.
From the rotation curve, corrected for asymmetric drift, we calculate the
co-rotation radius and find that the bar ends just inside this radius.
The two-dimensional character of these data allow us to show that the TW
method is sensitive to errors in the position angle (PA) of the disk.  
For example, a PA error of $2\degrees$ can give errors $\sim \pm 25\%$ 
in $\om$.
\end{abstract}


\section{Introduction}

Recent experience has demonstrated the utility of integral-field spectroscopy
to studies of galaxy dynamics.  Verolme \etal\ (2002), comparing mass models
of M32 constrained by integral-field spectroscopy from SAURON (Bacon \etal\ 
2001) and models constrained only by data along 4 slits extracted from the
full two-dimensional (2-D) data set, found that the mass model parameters,
including black hole mass, mass-to-light ratio and, in particular, inclination
were substantially better constrained with the full 2-D data than with just
the slits.

One approach to integral-field spectroscopy is the use of an imaging
Fabry-Perot interferometer (FP), essentially a tunable, narrow-band filter.
The Rutgers Fabry-Perot (RFP), formerly\footnote{The RFP has since been
  de-commissioned.  Replacement by the next generation RFP is scheduled for
  2004.} at CTIO\footnote{The Cerro Tololo Inter-American Observatory (CTIO)
  is part of The National Optical Astronomy Observatories (NOAO) which are
  operated by the Association of Universities for Research in Astronomy (AURA
  Inc.), under cooperative agreement with the National Science Foundation.},
found considerable application to the emission-line study of kinematics and
mass modeling of disk galaxies (\eg \ Buta \& Purcell 1998; Buta \etal\ 1999
\& 2001; Beauvais \& Bothun 1999 \& 2001; Palunas \& Williams 2000; Weiner
\etal\ 2001 a,b).  However, FP stellar absorption-line spectroscopy in
galaxies is significantly more difficult due to the lower surface brightness 
levels, the shallowness and width of the spectral features, and the effects of
time-variable observing conditions on reconstructing line profiles necessarily
sampled sequentially over an extended period of observation.  Because of these
difficulties, FP stellar absorption-line spectroscopy has not been previously
successfully attempted in galaxies, although absorption-line spectroscopy of
individual stars in globular clusters has been accomplished with considerable
success (Gebhardt \etal\ 1994, 1995 \& 1997).  In this paper, we study the
kinematics of a bright, barred, disk galaxy with stellar absorption-line
spectroscopy using the RFP, which is well-suited to galactic spectroscopy
because of its large field-of-view, high throughput, and appropriate spectral
resolution ($R = \lambda / \delta \lambda \simeq 2000$).

More than half of all high surface brightness disk galaxies are barred
(Knapen 1999; Eskridge \etal\ 2000).  The fundamental parameter in the
dynamics of barred galaxies is the pattern speed of the bar, $\om$.
This is usually expressed in terms of the quantity $\vpd = \lag/\len$;
here $\len$ is the length of the semi-major axis of the bar and $\lag$ is the
Lagrangian/corotation radius, where, in the bar's rest-frame, the
centrifugal and gravitational forces balance.  A self-consistent
bar with orbits aligned with the bar has $\vpd \geq 1$ (Contopoulos
1980); when $1.0 \leq \vpd \ltsim 1.4$, a bar is termed fast.

Because bars have strong quadrupole moments, they lose angular momentum
efficiently in the presence of a dense dark matter (DM) halo, as first shown
by perturbation theory calculations (Weinberg 1985).  Early simplified
simulations (Little \& Carlberg 1991, Hernquist \& Weinberg 1992) found
agreement with Weinberg's calculations.  The presence of DM in disk galaxies
is required by their rotation curves, which stay flat out to large radii
(Rubin \etal\ 1980; Bosma 1981).  Determining the relative amounts of luminous
and dark matter in the region of the optical disk, however, is a difficult
problem, due to degeneracy in the contributions of the two components to the
rotation of axi-symmetric systems.  The maximum disk hypothesis (van Albada \&
Sancisi 1986), which requires the maximum amount of disk matter consistent
with the inner rotation curve, often accounts for the shapes of the rotation
curves of high surface brightness (HSB) galaxies (\eg~Kalnajs 1983; Kent 1986;
Corsini \etal\ 1999; Palunas \& Williams 2000).  Others studies, however, have
argued that maximum disks cannot account for other observational properties
(\eg~van der Kruit 1995; Bottema 1997; Courteau \& Rix 1999, Kranz \etal\ 
2001).  Debattista \& Sellwood (1998) realized, from fully self-consistent
simulations, that fast bars require that barred galaxies have maximum discs.
At present, there is a lively debate about this suggestion (Tremaine \&
Ostriker 1999; Debattista \& Sellwood 2000; Weinberg \& Katz 2002; Valenzuela
\& Klypin 2003; O'Neill \& Dubinski 2003).

One of the issues surrounding this debate concerns just how rapidly bars
rotate.  The number of barred galaxies in which $\vpd$ has been measured
is still relatively small.  Evidence that bars are fast comes mostly
from simulations of the hydrodynamics of gas in the bar region
(\eg~Athanassoula 1992; Lindblad \etal\ 1996).  The example of NGC
4123 (Weiner \etal\ 2001b) is especially instructive because the
hydrodynamical simulations require both a fast bar and a maximum disk
to match the observed gas shocks.  A direct (model-independent) method
for measuring bar pattern speeds was devised by Tremaine \& Weinberg
(1984).  The Tremaine-Weinberg (TW) method is contained in the
following simple equation
\begin{equation}
\pin \om \sin i \ = \ \kin.
\label{eq:tweqn}
\end{equation}
Here, \mbox{$\pin = \int\,h(Y)\,X\,\Sigma\,dX\,dY$}, \mbox{$\kin =
  \int\,h(Y)\,V_{\rm los}\,\Sigma\,dX\,dY$}, $\Sigma$ is the surface
brightness, $h(Y)$ is an arbitrary weighting function, $V_{\rm los}$
is the line-of-sight velocity, $i$ is the disk inclination
(throughout, we use the convention $i = 0$ for face-on) and $(X,Y)$
are galaxy-centered coordinates measured along the disc's apparent
major and minor axes, respectively.  Although the integrals in $\pin$
and $\kin$ range over $ -\infty \leq X \leq \infty$, if the disk is
axisymmetric at $|X| \geq X_o$, then integration over $ -X_o \leq X
\leq X_o$ suffices.  The integration in $Y$ is over an arbitrary
range, because of $h(Y)$; for example a weight function proportional
to a delta function corresponds to a slit parallel to the disk major
axis.  For a number of such slits, plotting $\kin$ against $\pin$
gives a straight line with slope $\om \sin i$.  The TW method has
previously been applied to a handful of galaxies, all with slit data
(Merrifield \& Kuijken 1995; Gerssen \etal\ 1999; Debattista \etal\ 
2002; Aguerri \etal\ 2003; Gerssen 2003).  In all cases, bars were
found to be consistent with $1.0 \leq \vpd \ltsim 1.4$.  However, TW
measurements are technically challenging, and in some cases large
uncertainties result.  Thus, for example, Valenzuela \& Klypin (2003)
argue that $R = 1.7$, which they find in their simulations, is still
consistent with observations.  While correct, this claim overlooks the
fact that this limit is reached only in those galaxies with the
largest uncertainties (Aguerri \etal\ 2003, hereafter ADC03) and
ignores other sources of scatter (Debattista 2003, hereafter D03).
There exists, therefore, a need for at least a few high precision
measurements of $\vpd$.

Integral field velocity data permit TW measurements of $\om$ with
greater precision, because of the improved spatial coverage.  In this
paper, we present observations of \n7 using the RFP interferometer,
which allowed us to measure 2-D kinematics down to a surface
brightness $\mu_I \sim 21$ \mas.  In \S\ref{sec:n7079} we describe \n7,
emphasizing those properties which make it a suitable system for this
analysis.  In \S\ref{sec:phot}, we present photometry of \n7.  This is the
first use of the RFP for absorption-line spectroscopy of galaxies, and
therefore in \S\ref{sec:spect} we present the spectroscopy and reduction
methods.  We show that the resulting data are in good agreement with
data obtained by standard slit spectroscopy.  In \S\ref{sec:kinematics},
we study the kinematics of \n7\ and measure $\om$ with the TW method.
We show at this point that the measured $\om$ is quite sensitive to
small errors in the position angle (PA) of the disk.  Finally, in
\S\ref{sec:discussion}, we discuss our results and conclusions.

\section{\n7}
\label{sec:n7079}

In choosing candidates for observation, SB0 galaxies were selected
from various catalogues and examined in the {\it Digitized Sky
  Survey}\footnote{The {\it Digitized Sky Survey} was produced at the
  Space Telescope Science Institute (STScI) and funded by the National
  Aeronautics and Space Administration.}.  Candidates were chosen with
a bar at a position angle intermediate between the minor and major
axes of the disk (as required by the TW method), a number of bright
stars within $80\arcsec$ of the galaxy center (to permit determination
of temporal atmospheric transmission variations between the RFP
frames) and lacking strong spirals, patchy dust and significant
companions.  The galaxies needed to be bright and large, and have
known recession velocities (required for efficient use of the RFP).
Finally, we chose to use the CaII 8542.09 \AA\ absorption line because
it is quite strong, and the combination of the red color of SB0
galaxies and the moderately high CCD detector quantum efficiency in
this portion of the spectrum maximizes the measurement efficiency
here.  Since this region of the spectrum contains a host of night sky
emission features (see Osterbrock \etal\ 1996), we selected candidate
galaxies with red-shifts that placed the CaII line in relatively
emission-free regions.

\n7 (ESO 287-36) satisfied all these conditions.  \n7 is classified as
(L)SB(r)$0^0$ (Buta 1995), is large and bright ($2\farcm3 \times
1\farcm3$, $B_T = 12.46$ [de Vaucouleurs \etal\ 1991, hereafter RC3])
and has a measured recession velocity ($2670 \pm 25 $ \kms, [Lauberts \&
Valentijn 1989]).  The red-shifted wavelength of the CaII line is
therefore $8618$\AA; assuming a rotation velocity of about 260 \kms\ 
(Bettoni \& Galletta 1997) and allowing for the width of the CaII
stellar line and for the gradient in wavelength across the RFP frame,
we needed to scan the galaxy from $8608$\AA\ to $8631$\AA.  The sky
has no emission between $8598$\AA\ and $8613$\AA\ and some weak O$_2$
emission (which can be subtracted out) between $8613$\AA\ and
$8631$\AA.  Bettoni \& Galletta (1997) detected O[III] $\lambda =
5007$\AA\ emission extending out to $\sim 15\arcsec$, but this gas is
faintly emitting and is not expected to affect measurement of $\om$.
\n7 is the brightest member of a group containing seven galaxies
(Garcia 1993) but the nearest large neighbor, ESO 287-37, is over
$0\fdg5$ away from it, or $363$ kpc (in projection) at a distance of
38 Mpc (we assume throughout this paper that $H_0 = 70$ \kmsm).  The
other group members are more than twice as far away, so we expect that
\n7\ should not be significantly perturbed.

\section{Surface Photometry}
\label{sec:phot}

\subsection{Observations and Reduction}

\begin{figure}
\plottwo{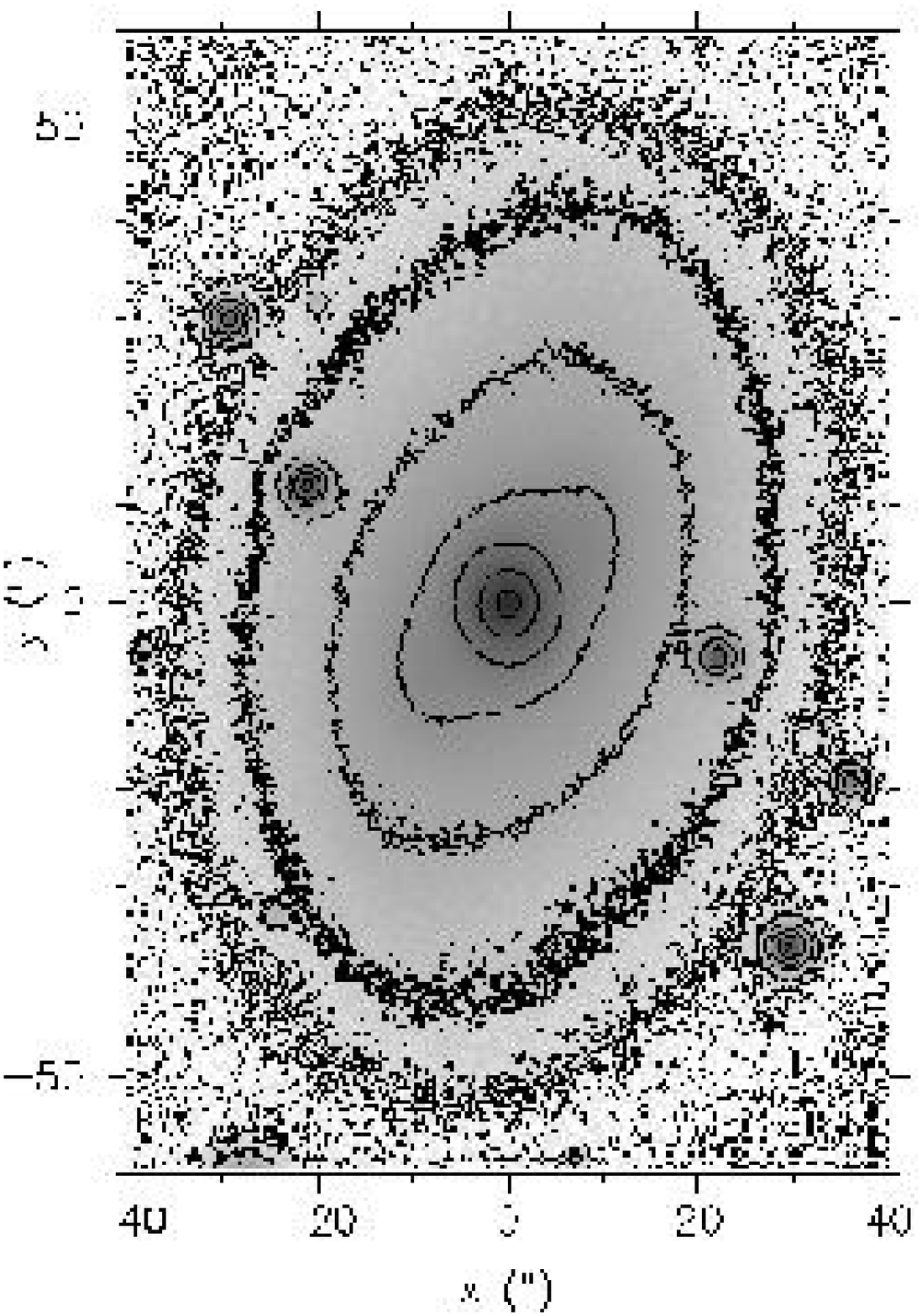}{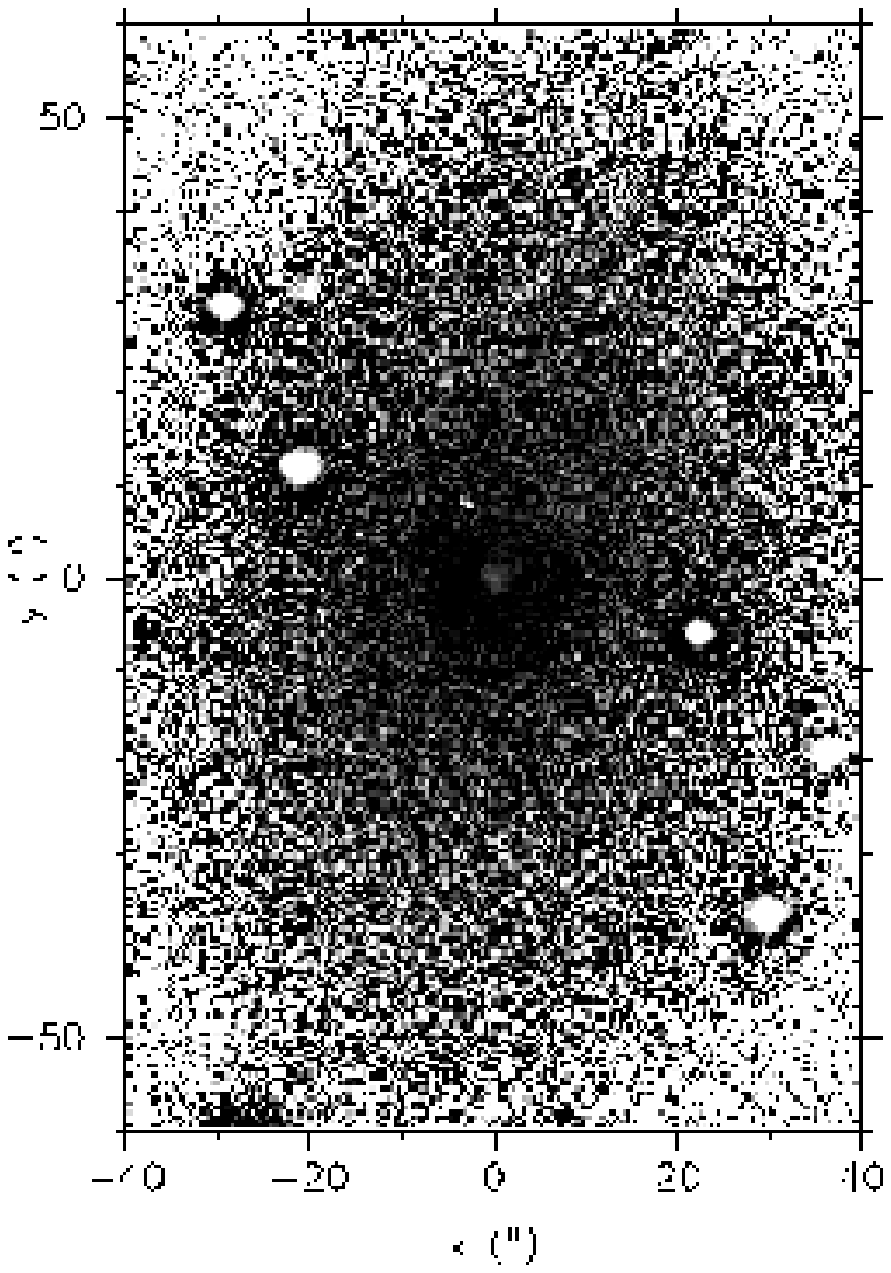}
\caption{The $I$-band (left panel) and the $B-I$ maps (right panel) of 
\n7.  North is left and west is up in each panel.  The bright stars near 
the galaxy were used to normalize the RFP frames.  The contours in the 
$I$-band map are spaced at 1 \mas, with the bold contour indicating
$\mu_I = 21$ \mas.  The only significant feature in the color map 
is the bulge at the center, with no evidence for a patchy obscuration.}
\label{fig:maps}
\end{figure}

We obtained deep images of \n7\ at the CTIO 0.9-m telescope on the night of
1997 August 2-3 under photometric conditions.  We used the TeK 2K\#3 CCD in
quad readout mode, with a gain of $3.3$e$^-$/ADU and read noise of $4.7$e$^-$
averaged over the four quadrants.  The image scale in this configuration was
$0\farcs396$/pixel, for an unvignetted field of $13\farcm5 \times 13\farcm5$.
The exposure time for each image of the galaxy was 900 seconds; we acquired
four exposures in $I$, two each in $B$, $V$, and $R$, and one exposure in $U$.
The seeing through the night was $\sim 1\farcs5$.  For the $I$-band images,
the telescope was moved between exposures so that the galaxy's image fell on
different parts of the CCD, to allow the effects of bad pixels to be removed
when the images were combined.  The images were flat-fielded with twilight sky
flats, and zero-subtracted as usual.  Using the \iraf\footnote{\iraf~is
  distributed by NOAO, which is operated by AURA Inc., under contract with the
  National Science Foundation} task {\sc imalign}, the $I$-band images were
shifted to the nearest pixel, and then combined using the biweight, a robust
statistical estimator of the mean that is insensitive to outliers (Beers et
al. 1990). (Because of the small pixel scale relative to the seeing, integer
pixel shifts were adequate, and avoided interpolation effects around image
artifacts.)  A constant sky value, obtained from image regions that were free
of stars and galaxy light, was subtracted from each image.  All were then
cleaned of cosmic rays using \iraf's task {\sc cosmicrays}.  The resulting
$I$-band image is shown in Fig.  \ref{fig:maps}.  A $B-I$ map, also shown in
Fig. \ref{fig:maps}, reveals little structure, apart from a nucleus bluer than
the disk, indicating that there is no variable obscuration within the galaxy
(as required for the TW method).

Two standard stars were used for calibration, E9-47-U and E1-44-S 
(Graham 1982).  We obtained the values of the Galactic extinction from
Schlegel \etal\ (1998) via NED\footnote{The NASA/IPAC Extragalactic
  Database (NED), is operated by the Jet Propulsion Laboratory,
  California Institute of Technology, under contract with the National
  Aeronautics and Space Administration.}. 
We did not measure the differential atmospheric extinction, but used
the standard value from Hamuy \etal\ (1992).  Since we observed \n7\ 
and the standard stars all at less than 1.1 air masses, the resulting 
uncertainties are insignificant for our purposes.

\subsection{Isophotal analysis}

\begin{figure}
\plotone{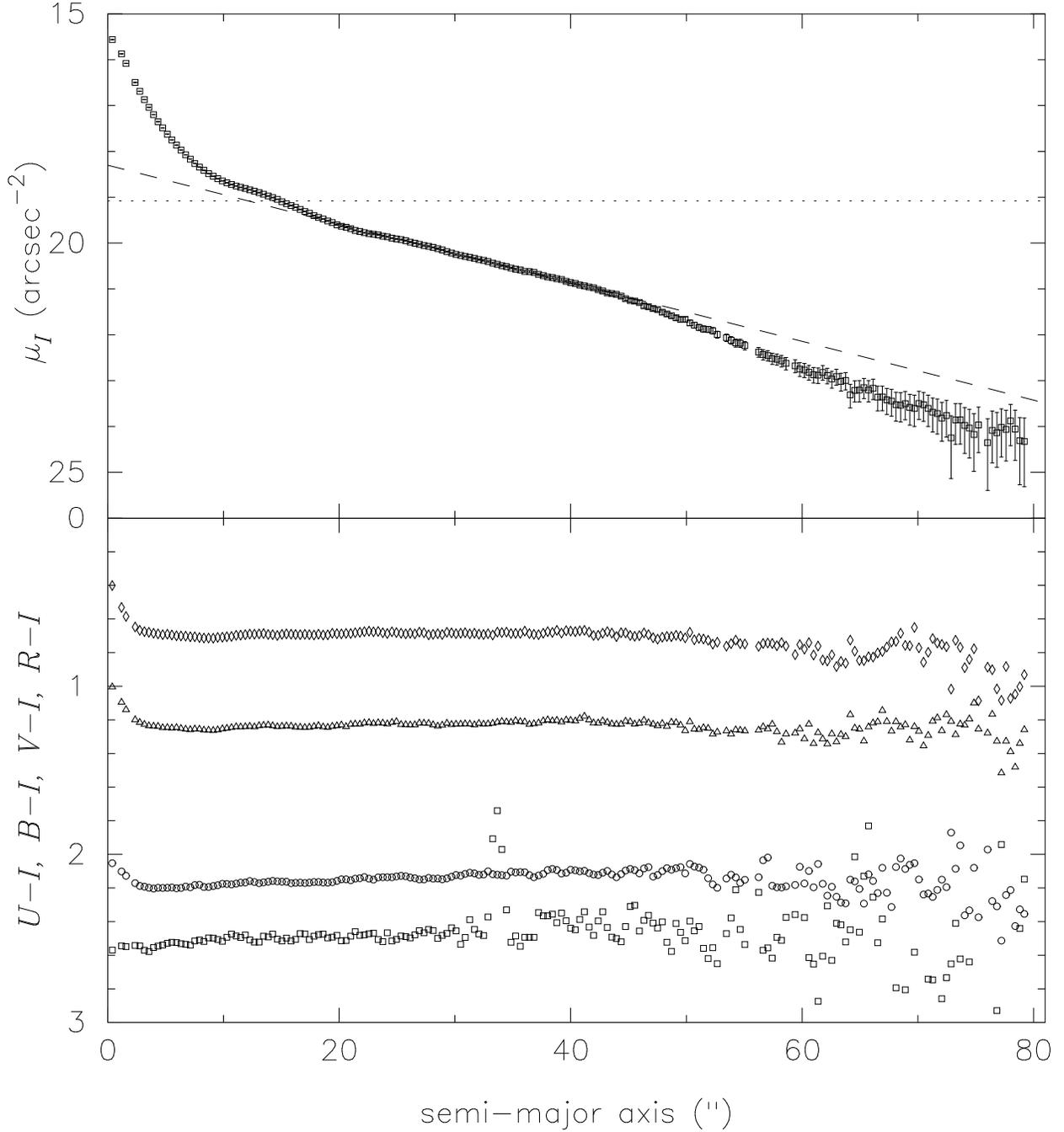}
\caption{The top panel shows the sky-plane surface brightness of \n7 in the 
  $I$-band.  The sky level is indicated by the dotted horizontal line,
  while the dashed line indicates the best-fit exponential within
  $25\arcsec < R < 40\arcsec$.  The radial profile exhibits a break at
  $\sim 45\arcsec$ ($\mu_I \simeq 21.2$ \mas).  The bottom panel shows
  the colors of \n7.  The squares are $U-I$, the circles $B-I$, the
  triangles $V-I$ and the diamonds $R-I$.  Note the quite flat colors
  (contamination by scattered light in the $U$-band is responsible for
  the feature seen in $U-I$ between $30\arcsec$ and $35\arcsec$).}
\label{fig:surfbright}
\end{figure}

After masking bad pixels, foreground stars, and all pixels outside
$100\arcsec$ from the galaxy center, we fitted ellipses, with fixed
centers, to single $U$, $B$, $V$ and $R$-band images, and to the
combined $I$-band image, using the \iraf~task {\sc ellipse}.  
We determined integrated magnitudes using the fitted ellipses and 
\iraf's task {\sc bmodel}.
We do not correct these magnitudes for the galaxy's inclination.   
Our results are reported in Table \ref{tab:magnitudes}, along with
the total asymptotic $U$, $B$ and $V$ aperture-magnitudes given in RC3.  
Our magnitudes for \n7\ are a bit fainter than RC3,
possibly because we masked out the bright stars in the vicinity
of the galaxy, and have not extrapolated the profiles to infinity.  To
measure the colors, we again used the task {\sc ellipse}, but now
constrained to measure the photometry in the same ellipses as were
fitted to the $I$-band, so that magnitudes can be meaningfully
compared in the same regions.  The results are plotted in Fig.
\ref{fig:surfbright}.  Apart from the slightly bluer nucleus, the color
profile is remarkably flat, with barely detectable color gradients.

We fitted separate exponential disks to the photometry in $B$, $V$, $R$ and
$I$ in the radial range $25\arcsec < R < 40\arcsec$, finding an average
scale-length, $R_d$, of $17\farcs1 \pm 2\farcs0$ (corresponding to $3.1 \pm
0.4$ kpc at our assumed distance) and $\mu_{0,B} = 20.5$ \mas.  The latter
value deprojects to 21.0 \mas, close to the Freeman (1970) limit ($21.65 \pm
0.3$ \mas).  At $R \gtsim 45\arcsec$, the surface brightness profile steepens.
This break radius remained unchanged when we fitted ellipses after changing
the constant sky background level by $\pm 1\sigma$, and is identical for $B$,
$V$, $R$ and $I$ bands, despite a difference in sky background level of three
magnitudes.  We conclude, therefore, that the break in the surface brightness
is not an artifact of imperfect sky subtraction.  Sharp disk truncations have
been noted in photographic plate studies (van der Kruit 1979), while in a
large sample of edge-on galaxies, Pohlen \etal\ (2002) found radial structure
that is better fit by two exponentials, rather than by sharply truncated
exponentials.  His break radii ($R_b/R_{\rm d} = 2.5 \pm 0.8$) and surface
brightnesses at $R_b$ ($\mu_R = 22.6 \pm 0.6$ \mas) agree well with the values
we obtain for the break in \n7 (2.6 and 22.0, respectively).  While the fact
that $R_b$ is identical in all bands argues forcefully that this is a real
structure, it is clear that the disk profile continues past $R_b$ (\cf\ 
Narayan \& Jog 2003).

\begin{figure}
\plotone{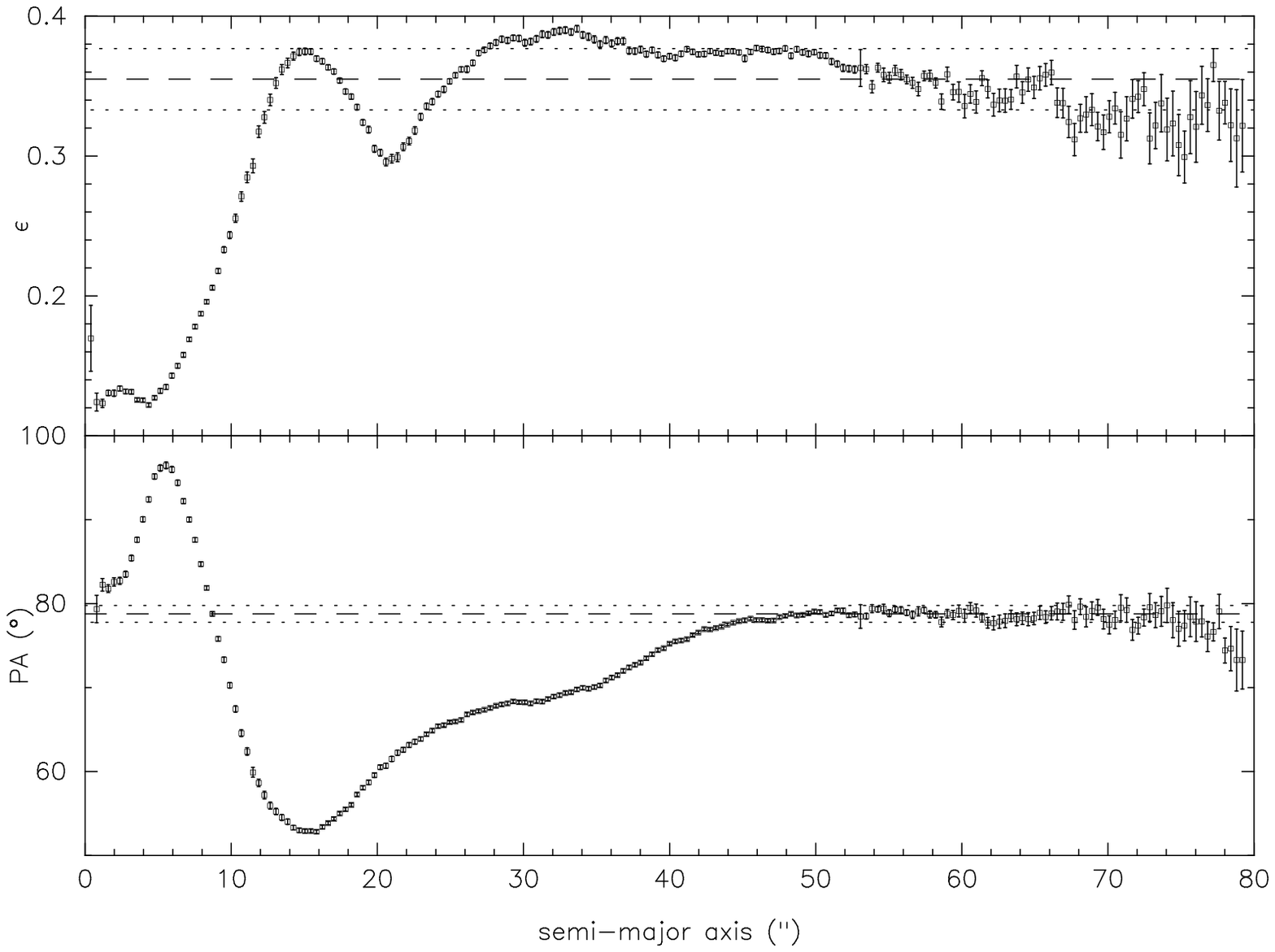}
\caption{The sky-plane ellipticity (top) and PA (bottom) of \n7.  The data 
from the ellipse fits to the $U$, $B$, $V$, $R$ and $I$ bands have been 
averaged together for this figure.  The disk's ellipticity and PA are 
determined from data with semi-major axis larger than $51\farcs5$.  This 
gives $\epsilon_{\rm disk} = 0.36\pm0.02$ and \padisk 
$~ =~ 78\fdg8 \pm0\fdg1$.  The average values of 
$\epsilon_{\rm disk}$ and \padisk\ are shown by the dashed lines; the standard 
error on $\epsilon_{\rm disk}$ is shown by the dotted lines, while the dotted
lines in the bottom panel show \padisk$~\pm~ 1\degrees$, illustrating that 
\padisk\ can be determined to better than $1 \degrees$ from these data.}
\label{fig:ellipses}
\end{figure}

We use data at semi-major axis larger than $51\farcs5$ to calculate the median
values of the disk ellipticity, $\epsilon_{\rm disk}$, and \padisk.  Fig.
\ref{fig:ellipses} shows these results averaged over all colors.  We find
$\epsilon_{\rm disk} = 0.36\pm0.02$ corresponding to an inclination $i =
49\fdg8 \pm 1\fdg7$ for a razor-thin disk.  If, however, the ratio of disk
scale height to scale length is $q_0 = 0.2$, we use the standard equation
(Hubble 1926) $\cos i = \sqrt{(b^2/a^2 - q_0^2)/(1-q_0^2)}$, to obtain
$i=51\fdg3$.  We therefore adopt $i = 50\fdg6 \pm 2\fdg3$, which is in good
agreement with RC3 ($i=51\fdg8$).  Our value of \padisk$ ~ = ~
78\fdg8\pm0\fdg1$; the RC3 value for this parameter is $82\degrees$.

\begin{figure}
\plotone{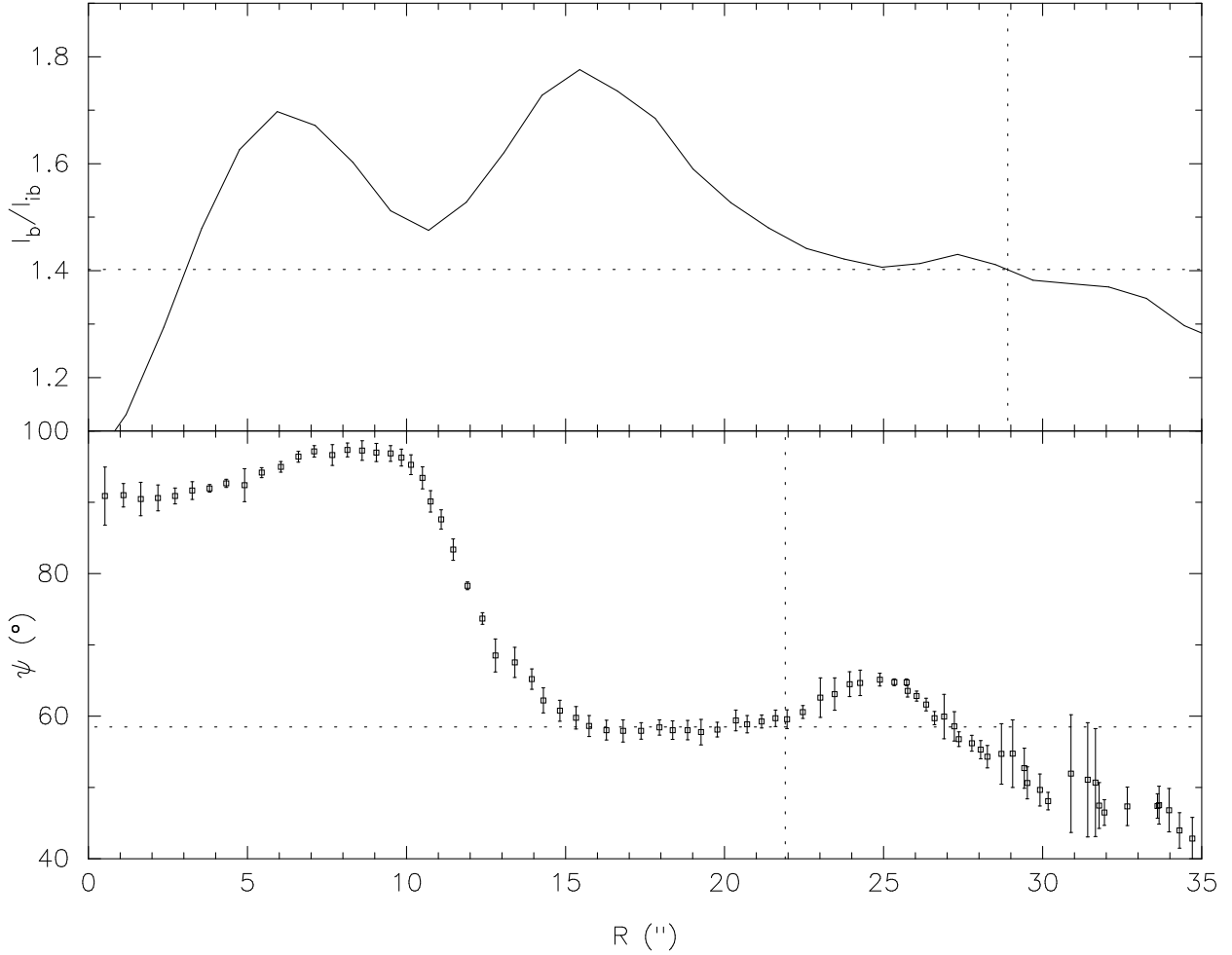}
\caption{The top panel shows the estimate of $\len$ based on the amplitudes
of the Fourier $m=2$ deprojected moment.  The dotted horizontal line
shows the value of $(I_b/I_{ib})_{\frac{1}{2}}$, from which 
$\len = 28\farcs9$, indicated by the 
vertical dotted line, is obtained.  The bottom panel presents the deprojected 
position angles relative to \padisk, $\psi$, of the ellipse fits of Fig. 
\ref{fig:ellipses} (averaged over all bands), in the inner part of the 
galaxy.  The vertical dotted line is at $\len = 21\farcs9$.  The horizontal 
dotted line shows $\psi_{\rm bar} \simeq 58\fdg5$.}
\label{fig:barlen}
\end{figure}

Since a bar represents an $m=2$ perturbation of the disk density, a Fourier
decomposition of the deprojected surface density is well-suited to measuring
the semi-major axis of the bar, $\len$.  Since two independent parameters, an
amplitude and a phase, define each $m \neq 0$ harmonic, two independent
estimates of $\len$ are possible from a Fourier decomposition.  Our first
estimate of $\len$, based on amplitudes, is given by the method of Aguerri
\etal\ (2000), which has been used in other early-type barred galaxies
(Debattista \etal\ 2002, ADC03).  The method defines $\len$ as the radius for
which \mbox{$(I_{b}/I_{ib})=(I_{b}/I_{ib})_{\frac{1}{2}} \equiv
  0.5[(I_{b}/I_{ib})_{max}+ (I_{b}/I_{ib})_{min}]$}, where $I_b = 1 + A_2$,
$I_{ib} = 1 - A_2$ and $A_2$ is the amplitude of the $m=2$ Fourier component,
normalized by that of $m=0$.  For \n7\, this prescription gives $\len =
28\farcs9$.  For a phase-based estimate of $\len$, we deprojected the ellipse
fits and measured the phase angle of the resulting ellipses.  The value of
$\len$ is then the largest radius out to which the phases are consistent with
a constant, taking into account that deprojecting the bulge, which we did not
subtract from our image for this analysis, results in a twist interior to the
bar.  We found $\len = 21\farcs9$ this way; we therefore adopt $\len =
25\farcs4 \pm 3\farcs5$.  Fig. \ref{fig:barlen} presents the bar-length
analysis and also shows that $\psi_{\rm bar}$, the bar-angle in the intrinsic
galaxy frame as measured from the major-axis of the disk, is $\simeq 58\fdg5$,
which projects to PA$_{\rm bar} =~ 32\fdg8$.

\section{Spectroscopy}
\label{sec:spect}

\subsection{Observations and Reduction}

We observed \n7 on 1997 July 28 with the CTIO 4-m telescope using the RFP
imaging interferometer and the TeK 2K\#6 CCD.  The instrument has a circular
field of view of radius $\sim 85\arcsec$, and the detector was operated with
$2 \times 2$ pixel binning, producing an image scale of $0\farcs70$/pixel.
The detector gain and read noise were $1.0 \pm 0.1$ e$^-$/ADU and $2.9 \pm
0.2$ e$^-$, respectively.  We used the ``broad'' RFP etalon, with a resolution
FWHM $\sim 4.2$ \AA\ (150 \kms) and free spectral range of $\sim 100$ \AA\ 
at the CaII
line.  The spectral instrumental profile of the RFP is well fitted by a Voigt
profile, the convolution of a Lorentzian and a Gaussian of widths $\sigma_l =
1.93$ \AA\ and $\sigma_g = 0.45$ \AA, respectively.  We used a blocking filter
of FWHM 100 \AA\ to select the desired spectral order.  The wavelength
calibration of the RFP was determined with a precision of $\sim 0.04$ \AA.
Because the wavelength zero point and the optical center drift during the
night (due to flexure, temperature changes, \etc), we took additional
calibration measurements hourly through the night and incorporated these
corrections in the data reductions.  We found average drift rates of $0.086$
\AA/hour and $0.36$ pixels/hour for the zero point and the optical center,
respectively.  The temperature dropped during the night, until the telescope
primary mirror was some $2.5\degrees$C warmer than the ambient air, producing
poor seeing of $1\farcs6-3\farcs0$ (FWHM).  There were also strong winds,
reaching 50 mph by morning and the night was non-photometric, with thin
variable cloud.  Nevertheless, we managed to obtain 24 exposures of 900
seconds each, spanning the wavelength range $8608 - 8631$ \AA\ in steps of
$\sim 1$\AA\ (35 \kms\ at 8618 \AA).  We first stepped through the full
wavelength range in steps of $\sim2$\AA, and then filled in the intermediate
wavelengths with a second scan (one wavelength sample was inadvertently
repeated).

The images were zero subtracted and flat-field corrected (with the median of 3
dome flats at each wavelength) in the the usual way.  Cosmic rays were then
cleaned by replacement using \iraf.  In an imaging FP, the transmitted
wavelength varies quadratically with the distance from the optical axis; for
the RFP at CaII the field is $\sim 7.5$ \AA\ bluer at the outer edge than on the
optical axis.  Thus the sky background in each frame varies radially due to
the spectral energy distribution of the night sky.  We offset \n7\ to the
right half of the field of view, so that we could determine the sky spectrum
from the left half of the field.  Because some faint galaxy-light from the
outer disk does reach to the left side of the frames, as well as a ghost image
from reflections off the CCD and FP, we developed an iterative scheme to
determine and subtract the sky.
Briefly, the initial estimate of the sky spectrum was determined from the left 
half of the image, and subtracted from the entire image; this approximately
sky-free image was used to estimate and remove the contributions of the outer 
galaxy and the ghost reflection from the left half of the original image, 
leading to an improved estimate of the sky spectrum.
A second iteration of this procedure produced the final sky spectrum that was
subtracted from the images.

The images were then masked, shifted to a common frame and convolved with a
radially symmetric Gaussian to a common seeing.  We did this twice, once for
all 24 images, (seeing FWHM $=2\farcs6$) and again for the best 18 images,
excluding those obtained under poor conditions at the end of the night
(resulting FWHM $= 2\farcs2$).  Since the atmospheric transparency varied by
$\sim 12\%$ during the night, all images needed to be normalized to a common
transparency.  We determined the normalization coefficients by measuring the
flux of seven bright stars in our field and assuming their mean spectrum over
our wavelength range was flat and featureless.  Finally, we generated noise
maps which include photon, sky and read noise, and which account for the
Gaussian and image registration smoothings.

To determine the kinematic properties of \n7, we fit a Voigt profile to the
spectral data at each spatial position in the FP data cube.  The data to be
fit were smoothed with box car averaging using an adaptive kernel ranging from
$1\times 1$ to $7\times 7$ pixels based on the flux level, and the fits were
weighted by the inverse square of the estimated flux uncertainties.  The
parameters of the fit were the continuum level, the line strength, the
line-of-sight velocity, and the Gaussian velocity dispersion; maps of these
quantities and their uncertainties were produced.  We were able to obtain
reliable fits within an ellipse of semi-major axis length of $\simeq
40\arcsec$ ($\mu_I \simeq\ 21$ \mas).  We did not attempt to fit higher
velocity moments, such as the Gauss-Hermite moments $h_3$ and $h_4$ (Gerhard
1993, van der Marel \& Franx 1993).  A comparison of the velocity fits
produced from the 18 and 24 image data cubes showed no significant
differences, mostly because of the poor quality of the last 6 images; in the
following, therefore, we consider only the fits based on the better-seeing (18
image) data cube.

\begin{figure}
\plottwo{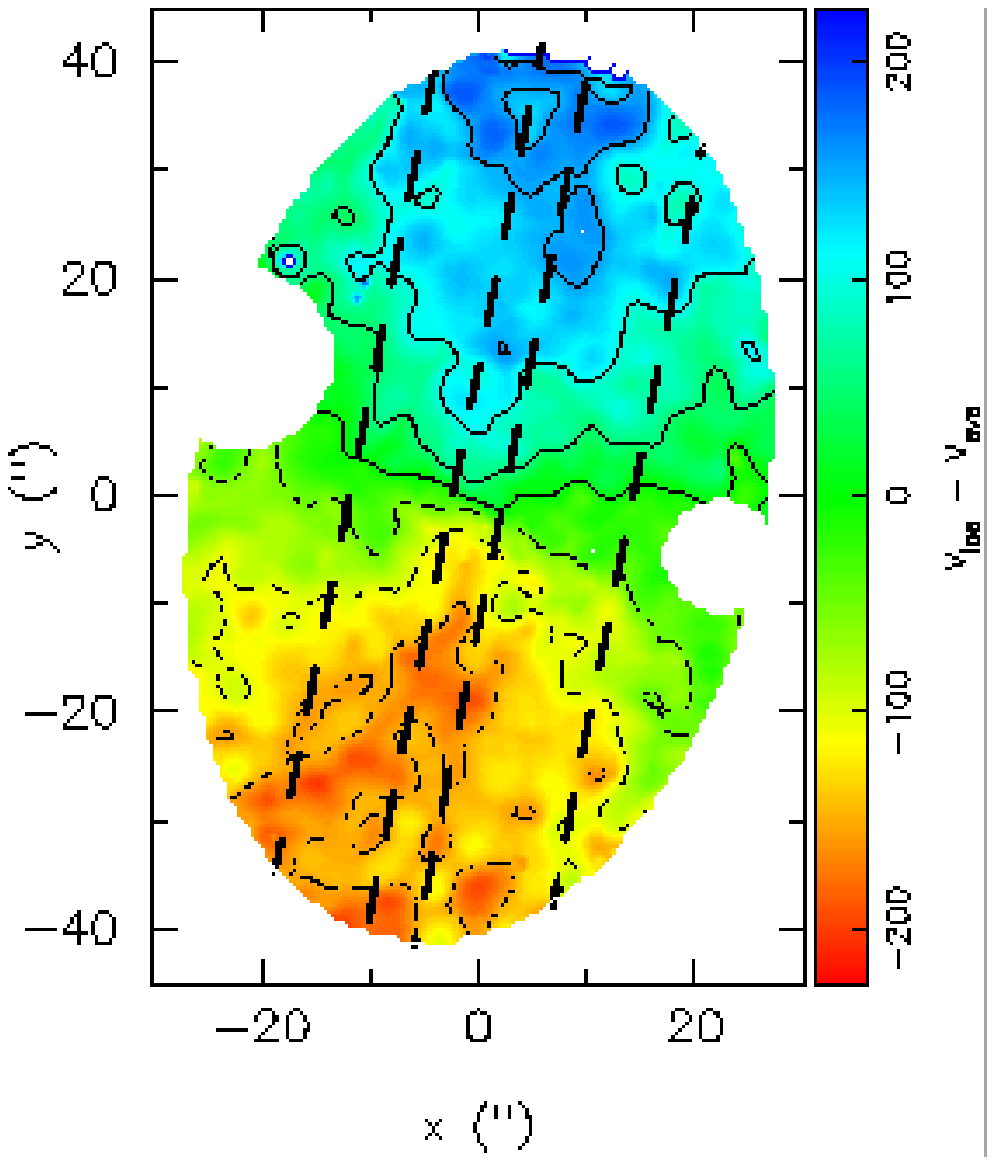}{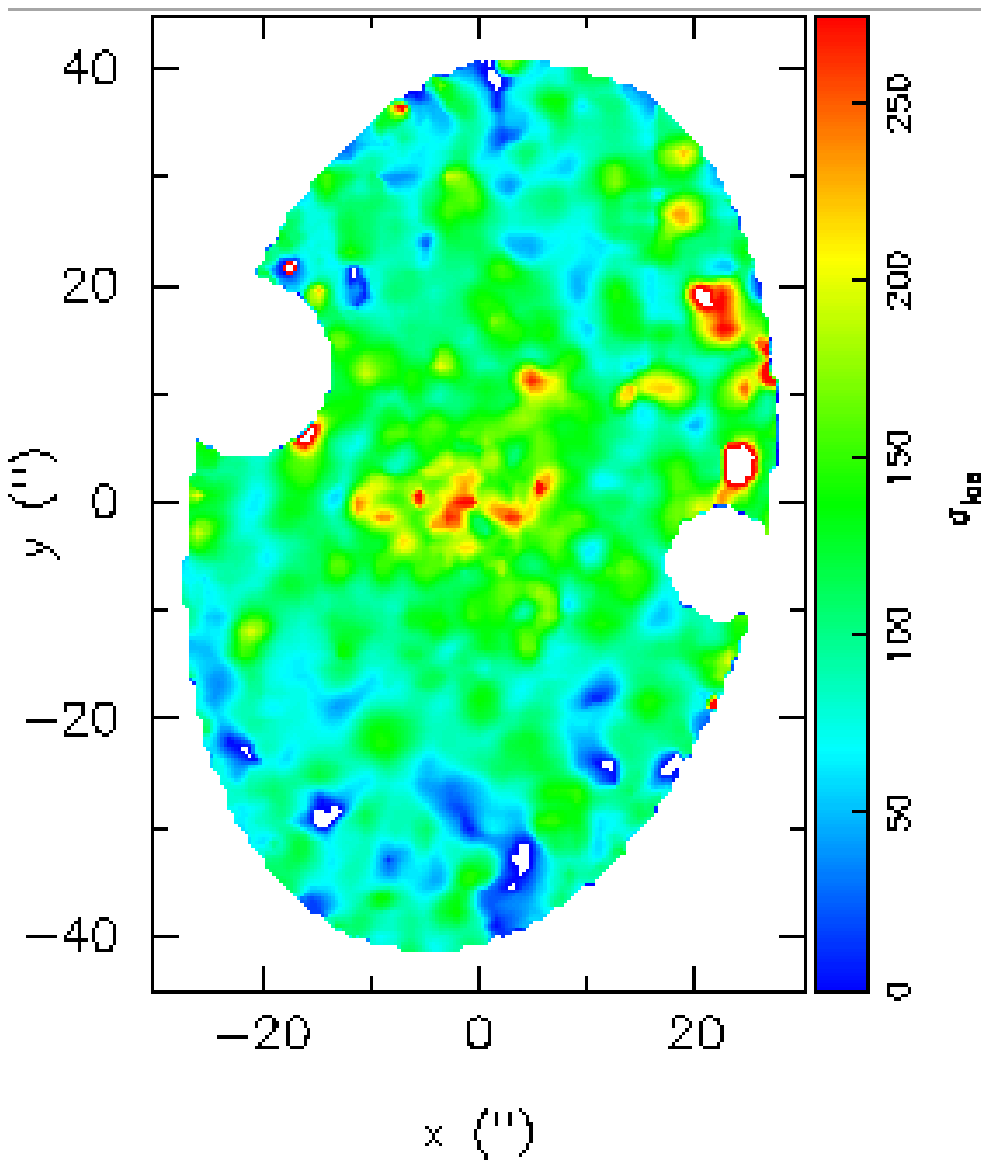}
\caption{The line-of-sight velocity (left) and velocity dispersion (right)
  fields; north is left and west is up.  In the velocity panel,
  contours are spaced by 50 \kms, centered on the systemic velocity
  and are solid (dashed) on the approaching (receding) side.  The bold
  dashed lines in the velocity panel indicate the 2 regions within
  which we applied the TW analysis.  Our limits on large $|Y|$ were
  chosen to avoid contamination by starlight, and the limits on small
  $|Y|$ are intended to avoid a small offset in the kinematic center.}
\label{fig:velsigfld}
\end{figure}

\begin{figure}
\epsscale{0.9}
\plotone{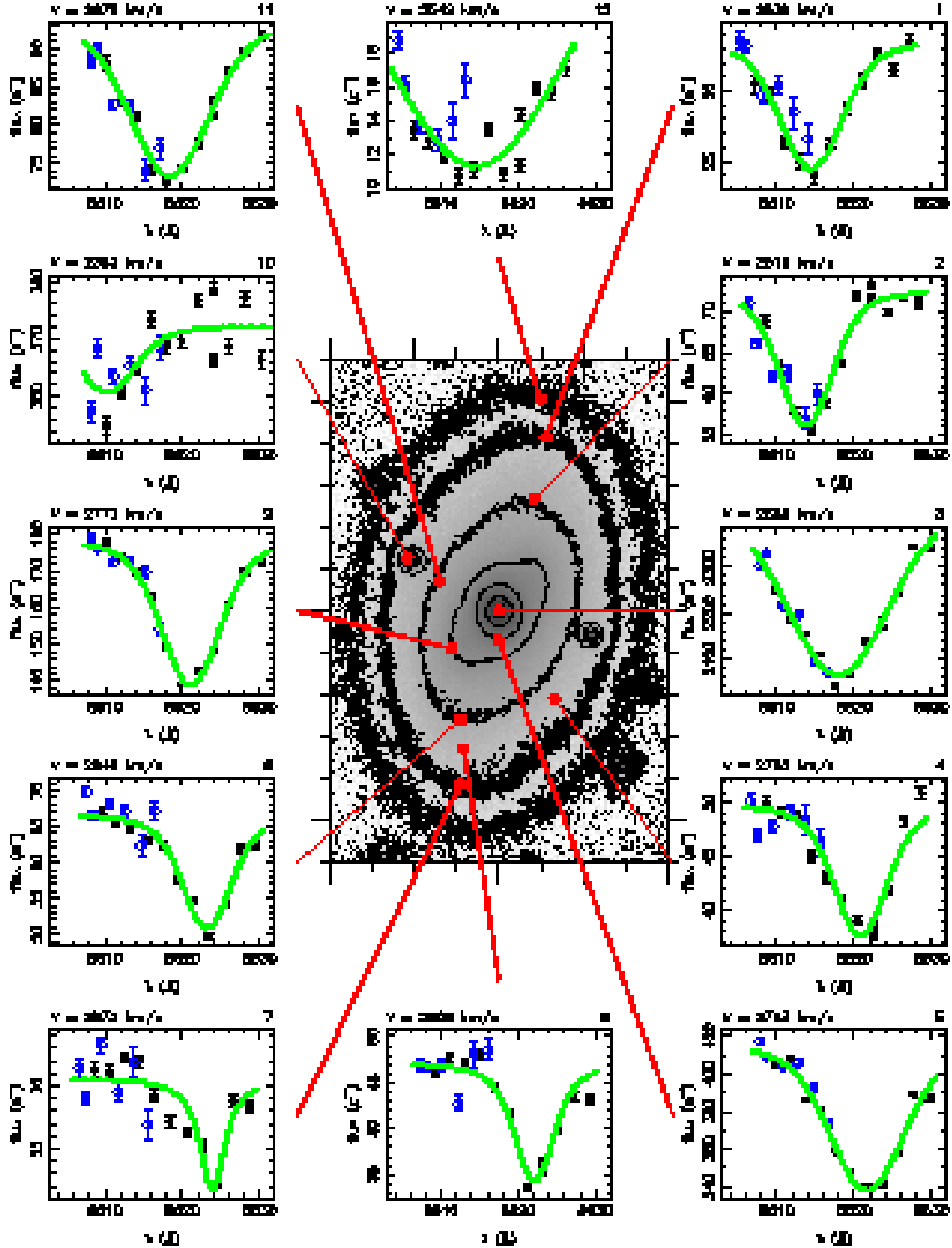}
\caption{Sample profile fits; each fit-panel (numbered in its 
upper-right and with the resulting $V_{\rm los}$ indicated in its 
upper-left) points to the location in the galaxy (here shown by the 
$I$-band map from Fig. \ref{fig:maps}) for which it applies.  The data
points indicate the fluxes (in electrons) in the reduced data cube, while the
green lines show our fitted profiles.  The points are coded by the 
time order in which they were observed: the black (filled) points are the 
first 12 observed, while the blue (open) points are the last 6 observed.}
\label{fig:velprofs}
\end{figure}

Fig. \ref{fig:velsigfld} presents the resulting velocity and velocity
dispersion maps.  Other than a slight S distortion in the central regions due
to the bar, the velocity distribution is that expected for an inclined,
rotating disk.  The velocity dispersion is largest in the nucleus and bulge,
lower along the bar, and smallest in the disk.  Examples of fitted profiles
are presented in Fig. \ref{fig:velprofs}; these show successful fits from the
galaxy center (panel 3) to $\mu_I = 21$ \mas\ (panels 1 and 7).  By $\mu_I =
22$ \mas, the signal-to-noise ratio is too low and only the noise is fit
(panel 12).  The spectrum of one of the foreground stars (panel 10) is
essentially constant (with possibly a very weak Fe $\lambda 8611$ \AA\ line).
As seen in panel 3, the FP scan did not extend far enough in wavelength to
sample the continuum adequately in the high velocity dispersion regions in the
center of the galaxy; thus while the velocities are well-determined here, the
velocity dispersion measurements have larger systematic uncertainties.

\subsection{Comparison with slit spectroscopy}

Since this is the first application of the RFP to absorption-line spectroscopy
of galaxies, it is important to confirm that our results agree with those
obtained by traditional slit spectroscopy.  For this purpose, E. M.  Corsini
kindly obtained and reduced for us a long-slit spectrum along the major axis
of \n7.  The 1800 second exposure was taken in 16 November 2001 using the ESO
New Technology Telescope (NTT) with EMMI in red medium-dispersion
spectroscopic mode and using the grating No. 6 with 1200 grooves mm$^{-1}$ in
first order with a $1\farcs0 \times 5\farcm5$ slit.  The detector was the No.
36 Tektronix TK2048 EB CCD with $2048\times2048$ $24 {\rm \mu m}$ pixels.  It
yielded a wavelength coverage between about 4840~\AA\ and 5490~\AA\ with a
reciprocal dispersion of 0.320 \AA\,pixel$^{-1}$.  The instrumental resolution
was $1.19$ \AA\ (FWHM) corresponding to $\sigma_{\it inst}\approx30$ \kms\ at
5170 \AA .  The spatial scale was $0\farcs270$ pixel$^{-1}$.

The spectrum was bias subtracted, flatfield corrected, cleaned of cosmic rays,
corrected for bad pixels and columns, and wavelength calibrated using standard
{\tt MIDAS}\footnote{{\tt MIDAS} is developed and maintained by the European
  Southern Observatory}, as described in ADC03.  The line-of-sight velocity,
velocity dispersion, and the higher order moments, were measured from the
galaxy absorption features present in the wavelength range centered on the Mg
line triplet ($\lambda\lambda\,5164,5173,5184$ \AA), using the Fourier
Correlation Quotient method (Bender 1990; Bender \etal\ 1994).

\begin{figure}
\epsscale{0.8}
\plotone{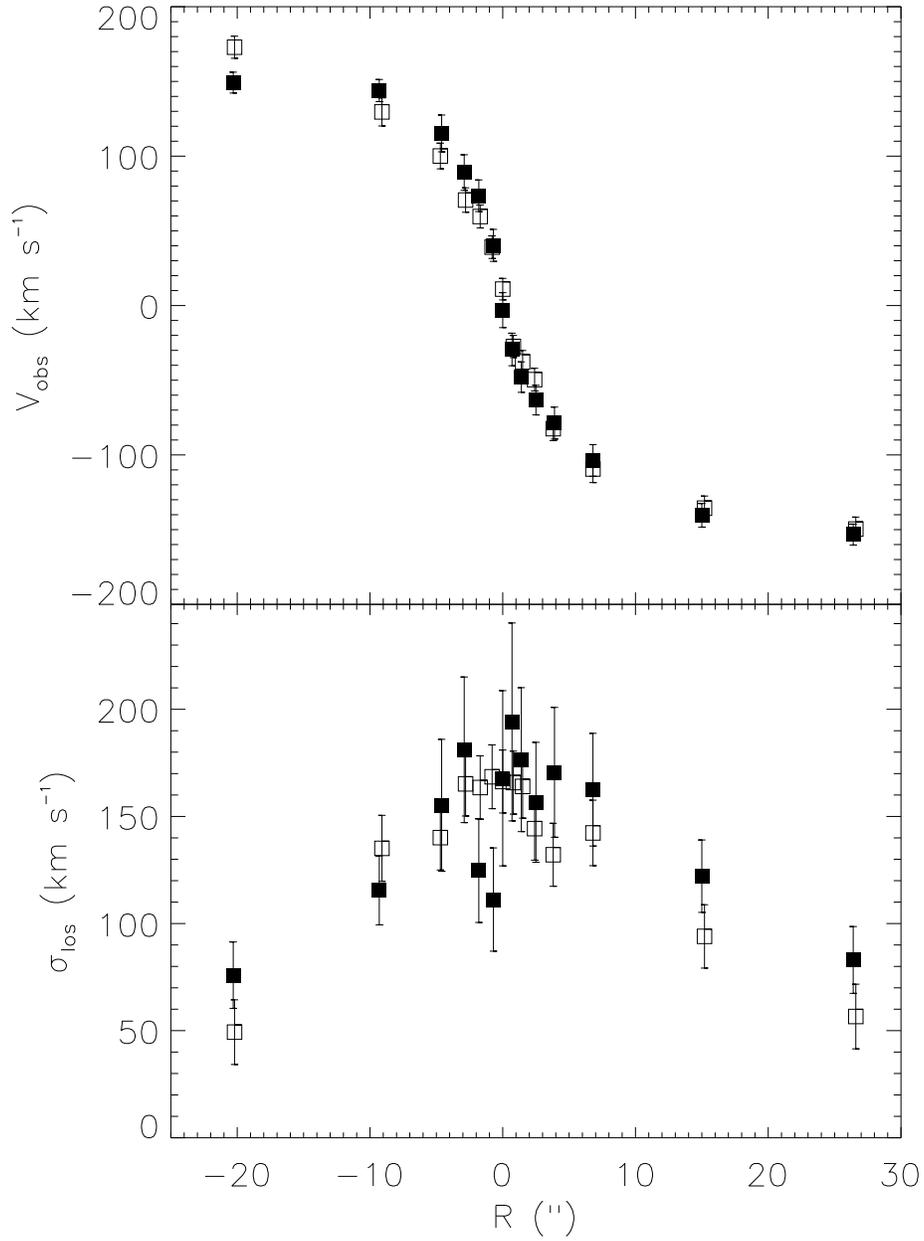}
\caption{A comparison of our data with slit data.  The top panel
  shows line-of-sight velocities, while the bottom one shows the
  velocity dispersions.  The open symbols are slit data, the solid symbols
  are the RFP data.  The error bars
  correspond to 1$\sigma$ deviations.}
\label{fig:slitcomp}
\end{figure}

In Fig. \ref{fig:slitcomp}, we compare the slit spectrum velocity and
dispersion data with data extracted from our RFP 2-D maps in a $1\farcs0$ wide
pseudo-slit along the major axis, smoothing along the pseudo-slit to match the
slit spectrum smoothing.  There is a zero-point offset of 14 \kms\ between the
two velocity data sets, which has been added to the plotted data.  The two
data sets agree well, with very similar amplitudes and shapes of the rotation
curves.  A chi-square analysis suggests that the velocity uncertainties are
under-estimated, and that an additional uncertainty of 8.4 \kms\ should be
added in quadrature to the error estimates.  If all this additional
uncertainty is assigned to the RFP data, the typical uncertainty of our
velocity measurements is 12 \kms; if the additional uncertainty is divided
evenly between the two data sets, the typical RFP errors are 10 \kms\ (and
typical slit spectrum errors are 8 \kms).  The RFP velocity dispersion data
have more scatter and larger estimated errors than the slit velocity
dispersions.  As noted above, we expect the lack of sufficient wavelength
coverage to affect the precision of the RFP dispersion measurements when the
dispersions are large, as in most of the points along the pseudo-slit.
Nonetheless, the two data sets are in good general agreement, tracing similar
amplitudes and distributions of the velocity dispersion.  Again, a chi-square
analysis suggests that the uncertainties are under-estimated, and that an
additional uncertainty of 20.1 \kms\ should be added in quadrature to the
error estimates.  If all this additional error is assigned to the RFP
measurements, their typical precision is 31 \kms; if the additional error is
split equally between the two data sets, the RFP uncertainties are 27 \kms\ 
and the slit spectra errors are 15 \kms.
There is some marginal indication that the velocity dispersions
measured with the RFP are systematically larger than with the slit
spectrograph.  The mean difference of all the measurements is 8 \kms,
and the mean difference of the measurements at radial distances
greater than $7\arcsec$ (where the RFP profiles are adequately
sampled) is 15 \kms; these biases are smaller than the estimated
uncertainties, and the spectral resolution of the RFP (corresponding
to an instrumental dispersion of 64 \kms) makes precision measurements
of dispersions below 50 \kms\ problematic.

Since the RFP and slit spectrograph data attain comparable results and
accuracies, it is straightforward to compare the relative efficiencies of the
two methods.  Since the spectral resolution of this spectrograph was about 
2.2 times better than that of the RFP, it should have been possible to widen 
the slit to $2\arcsec$ and still measure velocities with accuracies comparable 
to the RFP.  In an 1800 second exposure, the
spectrographic data reached to a limiting radius of $20\arcsec$, while the RFP
data extended to $40\arcsec$, about 1.3 magnitudes fainter.  The total RFP
exposure time was $18 \times 900$ seconds.  The sky and galaxy flux levels
were sufficiently high that both instruments were in the photon noise limit
regime.  Thus, taking into account the telescope apertures and the noise from 
the sky background, a single slit
spectrogram with $2\arcsec$ slit would reach the same surface brightness about
$3.3 \times$ faster than the RFP.  However, the RFP data extend out to $\pm
20\arcsec$ along the galaxy minor axis; to cover this same area would require
20 spectrographic exposures.  Thus the RFP is approximately $6 \times$ more
efficient than a slit spectrograph for measuring the two-dimensional velocity
field of this galaxy.  

We therefore conclude that the RFP data is in good agreement with the
slit spectrum data, and that the RFP technique affords a reliable means 
for obtaining absorption-line galaxy kinematics with significantly greater 
efficiency that standard slit spectroscopy.

\section{Kinematics and Bar Pattern Speed}
\label{sec:kinematics}

\subsection{The rotation curve}

\begin{figure}
\plotone{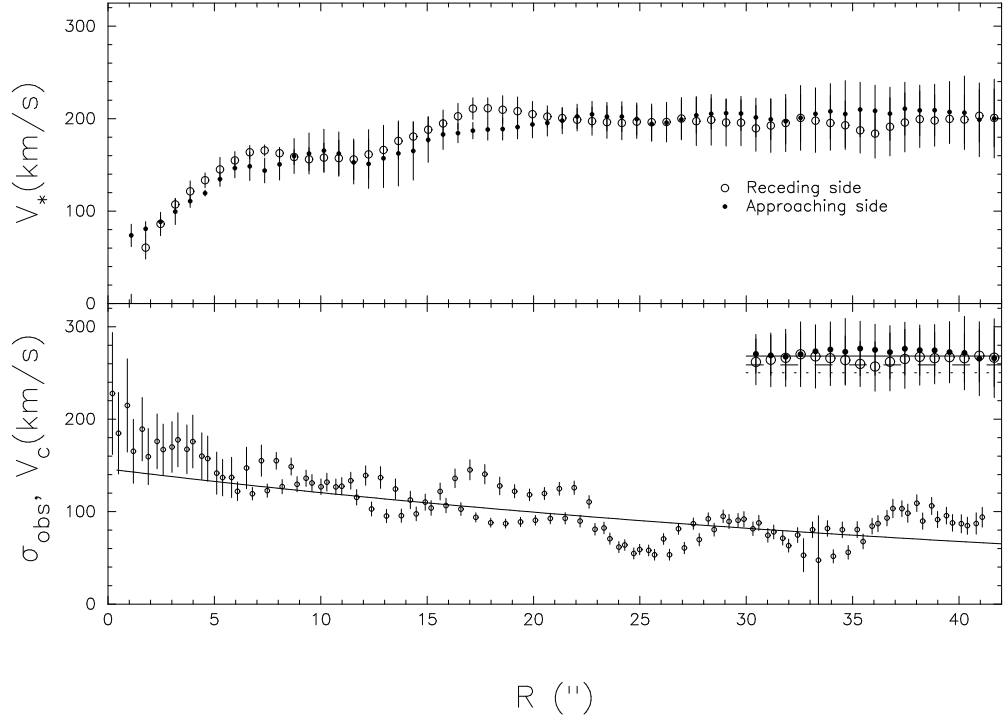}
\caption{The observed stellar streaming velocities, $V_{*}$, (top panel)
  and the asymmetric drift analysis for the amplitude of the rotation
  curve (bottom panel).  The bottom panel shows the observed velocity
  dispersion along the major axis of the disk; the best fit
  exponential is indicated by the solid line.  In the upper right
  corner of the lower panel are shown estimates of $V_{c,{\rm flat}}$
  (see Eqn. \ref{eqn:asymdrif}) for $\alpha = 0.7$ (closed and open
  circles) at $R \geq 30\arcsec$.  The radial average for this
  $\alpha$ is indicated by the solid horizontal line.  The dashed and
  dotted horizontal lines indicate the radial averages obtained by
  assuming $\alpha = 0.85$ and $\alpha = 1.0$, respectively.}
\label{fig:vels}
\end{figure}

Using the velocity field data, we obtained the stellar mean rotation
velocity curve, $V_*(R)$, by fitting tilted-rings (Begeman 1987,
Palunas 1996), with the galaxy center, inclination and \padisk\ fixed
to the values found in the photometry.  We started by determining the
systemic velocity, $V_{sys}$, obtaining $V_{sys} = 2670 \pm 4$ \kms.
This is in good agreement with Lauberts \& Valentijn (1989), who
obtained $V_{sys} = 2670 \pm 25$ \kms, and with the averaged values of
Bettoni \& Galletta (1997), $V_{sys} = 2678 \pm 35$ \kms.  This
parameter was then also fixed and $V_*(R)$ determined using pixels at
angle $|\phi| \leq 45 \degrees$ from the disk major axis in the
intrinsic galaxy frame.  The result is shown in Fig.  \ref{fig:vels}.
Subtracting $V_*(R)$ from the velocity map produced small residuals,
mostly $\ltsim 20$ \kms\ except near the bright foreground stars.

To determine the rotation curve of the galaxy we needed to correct for
the asymmetric drift, a procedure which is somewhat uncertain.  The
asymmetric drift equation for the circular velocity, $V_c$, is (\eg\ 
Binney \& Tremaine [1987] Eqn. 4-33)
\begin{equation}
V_c^2 - V_*^2 = -\sigma_R^2 \left[ {\frac{\partial \ln \rho}{\partial \ln R}} +
{\frac{\partial \ln \sigma_R^2}{\partial \ln R}} + \left( 1 - 
{\frac{\sigma_\phi^2}{\sigma_R^2}} \right) \right],
\label{eqn:asymdrif}
\end{equation}
where $\rho$ is the disk's volume density, and $\sigma_\phi$ and
$\sigma_R$ are the tangential and radial velocity dispersions in the
cylindrical coordinates of the galaxy's intrinsic plane.  To proceed
further, we needed to make a number of assumptions.  Following ADC03,
we assumed that the rotation curve is flat at large radii and that
both $\sigma_{\rm obs}$ and $\rho$ decrease exponentially with radius.
Thus the asymmetric drift equation becomes
\begin{equation}
V_{c,{\rm flat}}^2 = V_*^2 + \frac{[\sigma_0\exp(-R/R_{\sigma})]^2}
{\sin^2 i (1 + 2 \alpha^2 \cot^2i)}
\left[ 2 R \left( \frac{1}{R_{\rm d}} + \frac{2}{R_\sigma} \right) - 1 \right],
\end{equation}
where $\alpha = \sigma_z/\sigma_R$, $\sigma_0$ and $R_\sigma$ are the
parameters of the exponential fit to the observed major-axis velocity
dispersion profile, and $R_{\rm d}$ is the exponential scale-length of the
surface density.  As in ADC03, we use three values of $\alpha = 0.7, 0.85$ and
$1.0$, appropriate for early-type galaxies (Gerssen \etal\ 1997, 2000).

In Fig. \ref{fig:vels}, we present this asymmetric drift correction to
points at $R > 30\arcsec$.  We used $R_{\rm d} = 17\farcs1$ from the
photometry, as appropriate for the region of interest, and fitted an
exponential to $\sigma_{\rm obs}$ along a 7-pixel-wide slit on the
major axis.  We found a circular velocity amplitude, $V_{c,{\rm flat}}
= 256^{+17}_{-15}$ \kms.  The radially-averaged mean streaming
velocity in the same radial range is $\overline{V_*} = 200 \pm 5$
\kms.

\subsection{The pattern speed of \n7}

\begin{figure}
\plotone{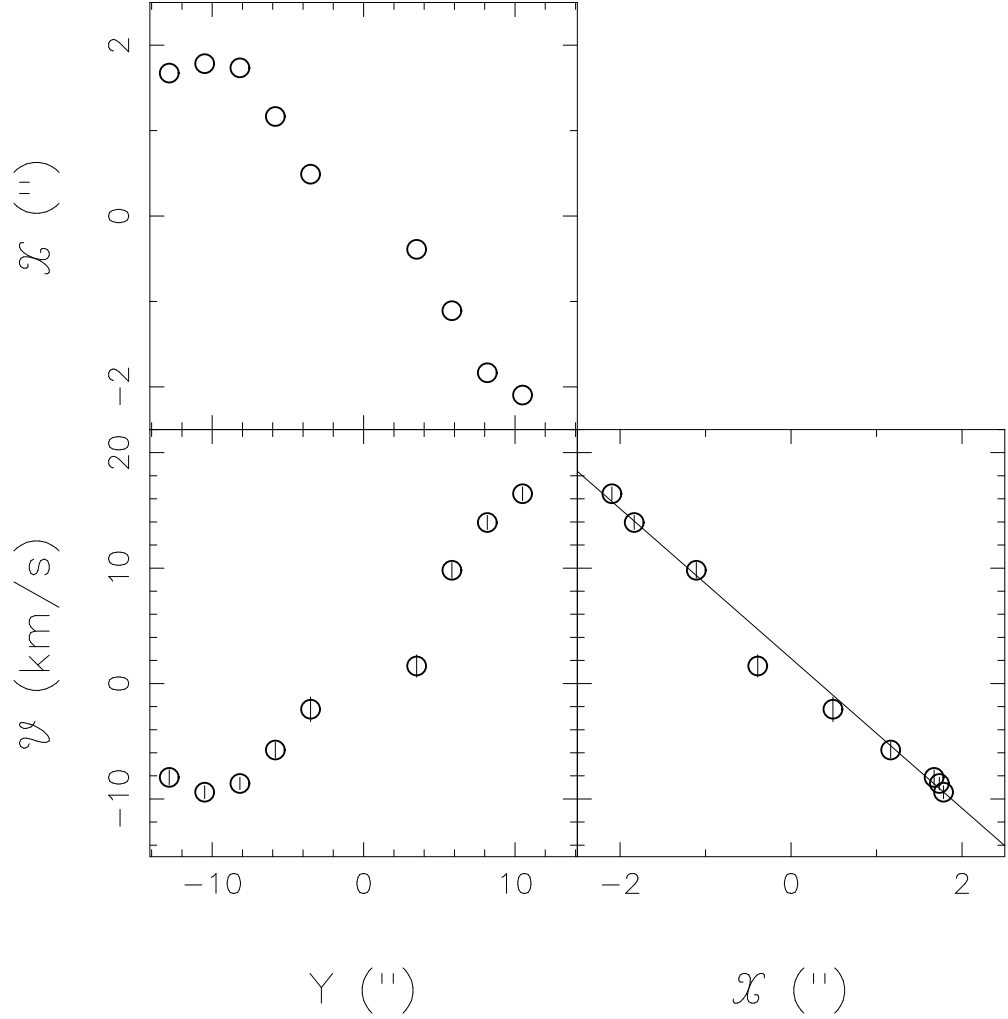}
\caption{The integrals $\pin$ (top left) and $\kin$ (bottom left) plotted as
  a function of $Y$.  The bottom right panel plots $\kin$ versus
  $\pin$; fitting a straight line gives slope $-6.5\pm 0.1$ \kmsa\ and
  $\tilde\chi^2 = 2.2$.  Excluding the most discrepant point reduces
  $\tilde\chi^2$ to 0.7 without changing the slope or its error
  significantly.}
\label{fig:twintegrals}
\end{figure}

We calculated the TW integrals for \n7\ in 9 strips in the ranges
$-14'' \leq Y \leq -2\farcs3$ and $2\farcs3 \leq Y \leq 11\farcs7$.
The limits on large $|Y|$ were chosen to avoid contamination by the
bright stars flanking \n7\ near its the minor axis.  The limits on
small $|Y|$ were chosen to avoid a small ($<1$ pixel) offset between
the kinematic and photometric centers, to which $\kin$ is sensitive
because of the centrally peaked light distribution.  In our standard
case, we calculated the integrals inside an ellipse of semi-major
axis, $A_{max} = 35\arcsec$, which does not quite reach the
axisymmetric part of the disk (see Fig.  \ref{fig:ellipses}), but
where the surface brightness is more than five magnitudes below the
central value (compare D03).  We chose the function $h(Y)$ for each
integral to be the integrated surface brightness within the strip.  We
obtained error estimates for $\kin$ from 100 Monte-Carlo realizations,
where the observed velocity field was modified at each point by adding
a velocity randomly drawn from a Gaussian distribution of width equal
to the estimated velocity uncertainty at that point plus an additional
8.4 \kms\ added in quadrature to account for the possible
underestimate of the velocity uncertainty.  In the standard case, we
used for $\Sigma(X,Y)$ the $I$-band map; to match the photometric and
kinematic maps, we first convolved the former to the seeing of the
latter, then used several of the stars in the field of both maps to
compute a geometric transformation between them.  This transformation
required only a translation and a single scale for $X$ and $Y$, with
no rotation or shear.  The results for the standard case are shown in
Fig.  \ref{fig:twintegrals}; the best fitting slope is $-6.5 \pm 0.1$
\kmsa\ (the negative sign here reflects only on our arbitrary choice
of positive $X$).

We checked that this value is not sensitive to various choices we made.  We
first tested that using the $I$-band photometry has not biased our result by
using instead an image generated by summing all the RFP frames.  The resulting
slope was $-6.3 \pm 0.3$ \kmsa, in good agreement with the standard case.  We
also checked that our limits on the integrals are reasonable by confirming
convergence of the values of $\pin$ and $\kin$ when we changed the semi-major
axis, $A_{\rm max}$, of the ellipse within which the integrals were performed.
In the range $ 31\farcs5 \leq A_{\rm max} \leq 38\farcs5$, the variations in
$\kin$ are all within the standard errors.  We use the variations in $\pin$ in
$ 35\arcsec \leq A_{\rm max} \leq 38\farcs5$ to define the errors in $\pin$;
these small errors lead to negligible change in the value of the fitted slope
and its uncertainty.

From this slope we calculate $\om = 8.4 \pm 0.2$ \kmsa, and $\lag =
V_{c,{\rm flat}}/\om = 30\farcs6_{-1.3}^{+1.4}$ (67\%confidence
interval).  We then find that $\vpd = V_{\rm c,flat}/(\len\om) = 1.2
\pm 0.1$ at the 67\% confidence interval and $\vpd =
1.2_{-0.2}^{+0.3}$ at the 99\% confidence interval, where we obtained
the uncertainties by Monte Carlo experiments in which we varied
$V_{\rm c,flat}$ and $\len$ uniformly in their error intervals and
$\om$ assuming Gaussian errors.  At a probability of higher than
$94\%$, \n7\ has a fast bar.  Note that, if the RFP over-estimates the
velocity dispersions in the region of interest, then the real $\vpd$
would be even lower; in the limiting (and unphysical) case that
velocity dispersions are negligible in the region of interest, we
obtain $\vpd = 0.9 \pm 0.1$.

\subsection{Variations in \padisk}

\begin{figure}
\plotone{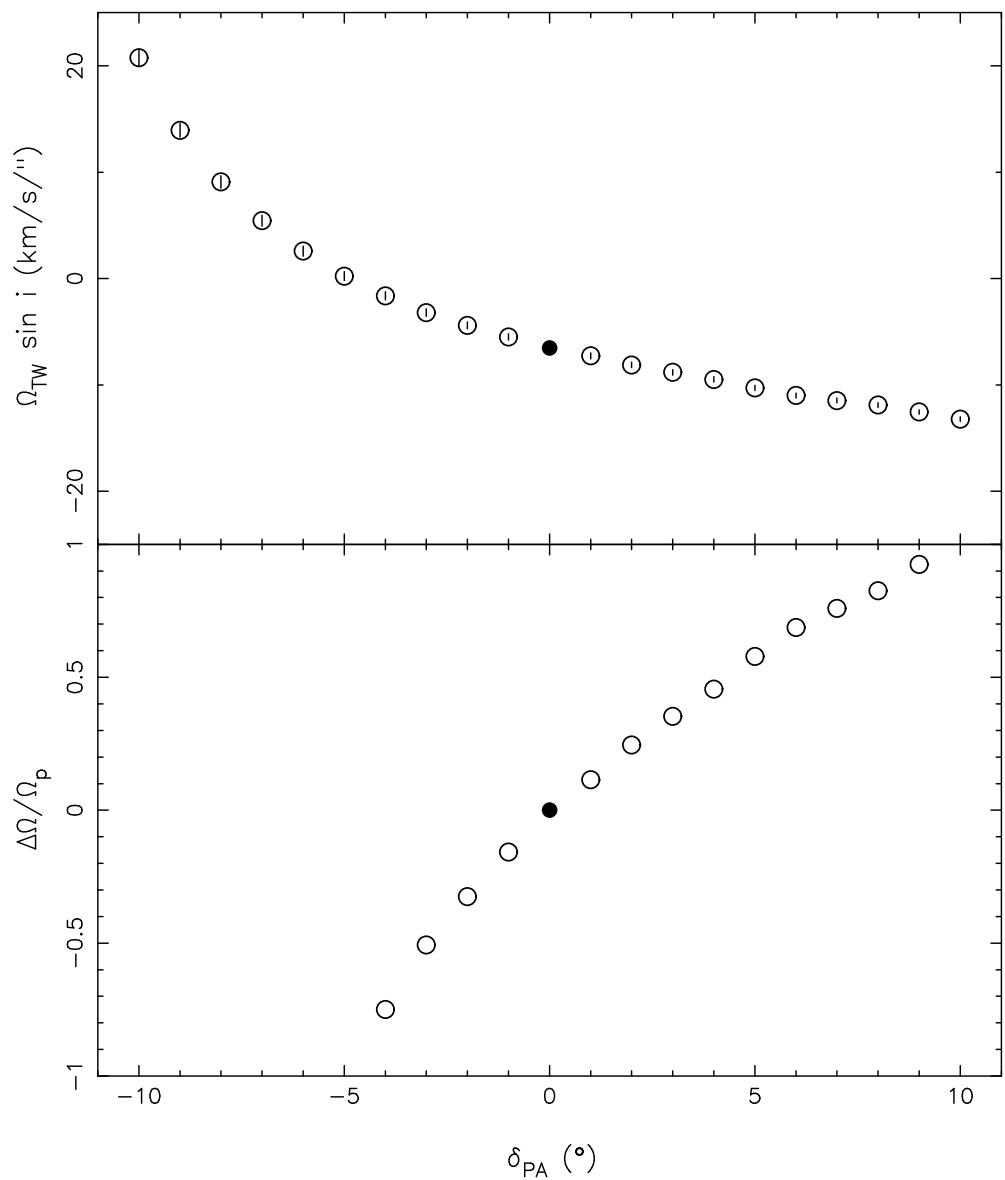}
\caption{The slope of the best fitting line (top panel) and the
  resulting error in $\Omega_{\rm TW}$, the pattern speed derived from
  the linear regression of the TW integrals (bottom panel), as
  functions of error in disk PA, $\dpa$.  In the bottom panel,
  $\Delta\Omega/\om \equiv (\Omega_{\rm TW} - \om)/\om$.  The solid
  circle in both panels represents the case when $\dpa = 0$.}
\label{fig:paerrs} 
\end{figure}

Our 2-D data also allowed us to study the effects of errors in \padisk\ on the
fitted slope.  In the convention where the error in \padisk, $\dpa$, moves
\padisk\ away from alignment with the bar when $\dpa$ is positive (the same
convention as in D03), we repeated our standard TW measurement as before but
for $-10\degrees \leq \dpa \leq 10\degrees$.  The results, plotted in Fig.
\ref{fig:paerrs}, reveal that the slope is quite sensitive to rather small
$\dpa$.  Errors in \padisk\ of order $5\degrees$, for example, produce large
errors in $\om$, $\sim 50\% - 100\%$.  Clearly, how accurately $\om$ can be
measured depends strongly on how well \padisk\ can be determined.  Our
high-precision determination of \padisk\ (see Fig. \ref{fig:ellipses}) allows
us to determine $\om$ with the small uncertainty quoted above.  Note however
that had we used the value of \padisk\ from RC3, our value of $\om$ would have
been some $35\%$ larger.

These results are consistent with those of D03 who studied $\Delta\Omega/\om$
systematically by means of an $N$-body simulation.  The sensitivity of $\om$
calculated here from the \n7\ observations is, if anything, somewhat larger
than in D03 (compare our Fig. \ref{fig:paerrs} with Fig. 9 in D03).  Because
\padisk\ errors can also be produced by disk density ellipticities, D03 was
able to show that such ellipticities had to be $\ltsim 0.07$ for the sample of
SB0 galaxies with TW measurements.

\section{Discussion and Conclusions}
\label{sec:discussion}

We have shown that FP techniques can measure galaxy kinematics from stellar
absorption-lines in early-type galaxies.  To the best of our knowledge, this
is the first successful demonstration of this technique.  Comparison with a
traditional slit spectrogram along the major axis of the galaxy confirms the
accuracy of the FP measurements.  For the galaxy observed here, mapping the
kinematics was about 6 times more efficient with the FP than with a slit
spectrograph.  For an object that fills more of the FP field of view, the
advantage would be even greater.  The fact that the spectral and spatial
resolutions are decoupled with an imaging FP affords a great deal of
flexibility in the data analysis, allowing extended measurements at lower
spatial resolution in the faint outer regions of galaxies.  Since the FP
samples a large area on the sky, the determination and subtraction of the sky
spectrum can be performed with high precision, again allowing measurement to
fainter surface brightnesses.  The full 2D format of FP kinematic maps permits
unambiguous determination of the positions at which velocity samples are
obtained, and allows optimal extraction of information in crowded fields,
which can be difficult with slit data.
 
There are some limitations to using a FP for absorption-line spectroscopy.
The temporal sampling of the spectrum limits the accuracy of reconstruction of
the line profile; this is the reason why we did not pursue measurement of
higher order moments of the velocity distribution function.  The next
generation RFP, with a built-in transparency monitor, should help alleviate
this difficulty.  The reduction of FP data has a reputation for difficulty; we
believe that this is more due to the wealth of information available in a full
2D kinematic dataset than to the inherent difficulty of analyzing the FP data.
Nonetheless, the techniques for handling FP data are not widely known, and FP
software is neither as generally available nor as robust as the packages for
slit spectroscopy.  Finally, while we have shown the significant efficiency of
a FP for producing a full 2D map of the stellar kinematics of a galaxy, there
are many classes of investigation for which such extensive data are not
required.  In such situations traditional slit spectroscopy is both simpler
and more efficient.

These problems should not, however, be overstated.  The promise of FP
absorption-line spectroscopy is demonstrated by the results obtained here.
This is the tightest bound on $\vpd$ ever obtained on a barred galaxy using
the TW method, with a 67\% interval less than half that for the next best
published case (ESO 139-G009, ADC03).  More dramatically, the fractional
uncertainty on $\om$ is less than $3\%$: typical uncertainties from slit
spectroscopy are $\sim 20-30\%$.

We showed that, in order to measure $\om$ to such high precision, the PA of
the disk needs to be known quite accurately.  Our's is the first demonstration
on a real galaxy that the TW method is sensitive to small errors in PA.  For
\n7, we were able to measure \padisk\ to better than $1\degrees$.  But if we
had relied on the RC3 value of \padisk\ we would have obtained a value of $\om
\sim 35\%$ larger.

We found, at high confidence, a fast bar in \n7.  As in all previous
measurements with good precision, the probability that $\vpd = 1.7$ is low: if
we assume extreme values for $\len$ and $V_{c,{\rm flat}}$ to maximize $\vpd$,
we find that $\vpd = 1.7$ requires $\om = 7.3$ \kmsa, or over $5\sigma$
smaller than our measured value.  Such fast bars continue to be a challenge for
cuspy cold dark matter halos.

\acknowledgments

We thank the CTIO staff for their excellent support and for trouble-shooting
instrument problems during the observing run.  We thank Povilas Palunas for
providing us with his tilted-rings velocity-field fitter.  We especially would
like to thank Enrico Maria Corsini for the spectrum and reduced kinematics
along the major axis.  V.P.D.  thanks J. A. L.  Aguerri for useful discussions
and cross-checks.  V.P.D.  acknowledges support by NSF grant AST 96/17088 and
the Swiss National Science Foundation Grants No. 20-56888.99 and 20-64856.01.
TBW acknowledges support by NSF grants AST-9731052 and AST-0098650.  The RFP
was developed with support from Rutgers University and the NSF, under Grant
AST-8319344.  We dedicate this paper to the memory of Bob Schommer, who was
instrumental in developing the RFP, turning it into a user instrument at CTIO,
and producing much excellent science with it.

\clearpage

\clearpage

\clearpage

\clearpage

\clearpage

\clearpage

\clearpage

\clearpage

\clearpage

\clearpage

\clearpage
\begin{deluxetable}{ccc}
\tabletypesize{\scriptsize}
\tablecaption{Apparent magnitudes of \n7. \label{tab:magnitudes}}
\tablewidth{0pt}
\tablehead{
\colhead{Filter} & \colhead{$m$} & \colhead{$m_T$\tablenotemark{a}} 
}
\startdata   
$U$ & 13.2 & $12.83 \pm 0.13$  \\
$B$ & 12.6 & $12.46 \pm 0.13$  \\
$V$ & 11.7 & $11.59 \pm 0.13$  \\
$R$ & 11.1 & \nodata           \\
$I$ & 10.4 & \nodata           \\
\enddata   
\tablenotetext{a}{RC3 (via NED)}
\tablecomments{We estimate that our uncertainty is less than
0.1 magnitude in all bands.}
\end{deluxetable}

\end{document}